\documentclass[preprint,superscriptaddress,nofootinbib]{revtex4-1}
\usepackage{graphicx}
\usepackage{subcaption}
\usepackage{float}
\usepackage{amsmath}
\usepackage{amssymb}
\usepackage[utf8]{inputenc}
\usepackage{hyperref}
\usepackage{color}

\begin{document}

\title{Search for the vectorlike leptons in the U(1)$_X$ model inspired by the $B$-meson decay anomalies}

\author{Fang-Zhou Xu}
\email{xfz14@mails.tsinghua.edu.cn}
\affiliation{Institute of Modern Physics, Tsinghua University, Beijing 100084, China}
\affiliation{CAS Key Laboratory of Theoretical Physics, Institute of Theoretical Physics, Chinese Academy of Sciences, Beijing 100190, China}
\author{Wenxing Zhang}
\email{zhangwenxing@itp.ac.cn}
\affiliation{CAS Key Laboratory of Theoretical Physics, Institute of Theoretical Physics, Chinese Academy of Sciences, Beijing 100190, China}
\affiliation{School of Physical Sciences, University of Chinese Academy of Sciences, No. 19A Yuquan Road, Beijing 100049, China}
\author{Jinmian Li}
\email{jmli@kias.re.kr}
\affiliation{College of Physical Science and Technology, Sichuan University, Chengdu, Sichuan 610065, China}
\affiliation{School of Physics, Korea Institute for Advanced Study, Seoul 130-722, Korea}
\author{Tianjun Li}
\email{tli@itp.ac.cn}
\affiliation{CAS Key Laboratory of Theoretical Physics, Institute of Theoretical Physics, Chinese Academy of Sciences, Beijing 100190, China}
\affiliation{School of Physical Sciences, University of Chinese Academy of Sciences, No. 19A Yuquan Road, Beijing 100049, China}

\begin{abstract}

We consider the U(1)$_X$ model, which can induce the flavor violating couplings through vectorlike fermions and address the observed rare $B$-meson decay anomalies.
To be consistent with all the other observations, both the associated gauge boson mass and the vectorlike lepton mass are bounded from above.
We argue that the search for new vectorlike leptons is promising and provides a complement to the $Z'$ search.
A detailed collider analysis shows that the model with the vectorlike lepton mass up to 1000 GeV could be tested at the future LHC.

\end{abstract}
\maketitle

\section{Introduction}

Since the LHCb collaboration observed a discrepancy with the standard model (SM) in the angular distribution of $B \to K^*(\to K \pi)\mu^+\mu^-$ \cite{Aaij:2013qta} in 2013,
$B$-meson anomalies have gained ever-increasing attention in the community. Benefitting from smaller hadronic uncertainties, the semileptonic $B$-meson decays provide clean probes of physics beyond the SM.
Over the past few years, the measured branching ratios \cite{Aaij:2014pli,Aaij:2016flj,Aaij:2015esa} and angular distribution observables \cite{Aaij:2015oid,Wehle:2016yoi} of rare $B$-meson decays induced by flavor-changing neutral-current transitions $b \to s\ell\ell$ are consistently in tension with the SM predictions.
Among these observables, the most clean ones are lepton flavor universality (LFU) ratios
\begin{equation}
R_{K^{(*)}}\equiv\frac{\text{BR}(B\to K^{(*)}\mu^+\mu^-)}{\text{BR}(B\to K^{(*)}e^+e^-)}~,
\end{equation}
where the hadronic form factors and potential systematic uncertainties cancel to a large extent.
Current data on $R_{K^{(*)}}$ \cite{Aaij:2014ora,Aaij:2017vbb} lie significantly below the SM predictions which are essentially unity \cite{Bordone:2016gaq}. For specified regions of the dilepton invariant mass squared,
\begin{equation}
R_K=0.745^{+0.090}_{-0.074}\pm0.036, \quad  1~\text{GeV}^2<q^2<6~\text{GeV}^2~,
\end{equation}
\begin{equation}
R_{K^*}=\left\{\begin{aligned}
0.66^{+0.11}_{-0.07}\pm0.03~,& &0.045~\text{GeV}^2<q^2<1.1~\text{GeV}^2\\
0.69^{+0.11}_{-0.07}\pm0.05~,& &1.1~\text{GeV}^2<q^2<6.0~\text{GeV}^2~.
\end{aligned}
\right.
\end{equation}

A heavy $Z'$ boson with flavor-changing couplings to quarks and nonuniversal couplings to leptons \cite{Altmannshofer:2013foa,Altmannshofer:2014cfa,Crivellin:2015mga,Ko:2017yrd,Crivellin:2015lwa,Celis:2015ara,Alonso:2017bff,Alonso:2017uky,Guadagnoli:2018ojc,Duan:2018akc,Gauld:2013qja,Buras:2013dea,Niehoff:2015bfa,Carmona:2015ena,Carmona:2017fsn,Gauld:2013qba,Buras:2013qja,Crivellin:2015era,Becirevic:2016zri,GarciaGarcia:2016nvr,Altmannshofer:2014cfa,Crivellin:2015mga,Ko:2017yrd,Sierra:2015fma,Allanach:2015gkd,Belanger:2015nma,Kim:2016bdu,Dalchenko:2017shg,Chiang:2017hlj,Chivukula:2017qsi,Allanach:2017bta} is an obvious candidate contributing to $b \to s$ anomalies.
In the literature, the $Z'$ boson is either associated with a horizontal gauge symmetry \cite{Altmannshofer:2014cfa,Crivellin:2015mga,Ko:2017yrd,Crivellin:2015lwa,Celis:2015ara,Alonso:2017bff,Alonso:2017uky,Guadagnoli:2018ojc,Duan:2018akc}, embedded in the 3-3-1 model \cite{Gauld:2013qja,Buras:2013dea}, the composite Higgs model \cite{Niehoff:2015bfa,Carmona:2015ena,Carmona:2017fsn}, or has generic couplings to quarks and leptons \cite{Gauld:2013qba,Buras:2013qja,Crivellin:2015era,Becirevic:2016zri}. Some models also employ vectorlike particles to generate required couplings \cite{Altmannshofer:2014cfa,Crivellin:2015mga,Ko:2017yrd,Sierra:2015fma,Allanach:2015gkd,Belanger:2015nma}.
In this paper, we consider the U(1)$_X$ model defined in Ref. \cite{Sierra:2015fma}, which introduces one generation of vectorlike quarks, one generation of vectorlike leptons and two complex scalar fields with one of them being a dark matter candidate.
Through mixings of the U(1)$_X$ charged vectorlike particles with their SM counterparts, LFU is violated in a manner similar to the SM with flavor-dependent Yukawa couplings. 
The model solves the $b \to s$ anomalies while providing a dark matter candidate and can be extended to generate neutrino masses. They conducted a numerical analysis of the Wilson coefficients and the dark matter relic density with respect to $g_X$ and $m_{Z'}$. Since we focus on collider searches instead of dark matter phenomenology in this work, only one complex scalar field is considered and the masses and Yukawa couplings of the vectorlike fermions are presumed to be varying rather than fixed.

There have been a number of works on studying the collider phenomenology of possible new physics (NP) that could address the $B$ anomalies, many of which \cite{Kim:2016bdu,Dalchenko:2017shg,Chiang:2017hlj,Chivukula:2017qsi,Allanach:2017bta} focus on the signature of $Z'$ decaying into dimuon.
In our setup, if we impose perturbativity requirements on the relevant gauge and Yukawa couplings, then $B_s-\bar B_s$ mixing measurements would imply an upper bound on the vectorlike lepton mass and a tighter constraint on the $Z'$ mass. 
The prospect of detecting these two particles at a future collider will be discussed in detail. Searches for them are complementary to each other in some regions of parameter space. Their combined sensitivity at the high luminosity LHC will be able to cover a broader range of parameter space that can explain the $B$ anomalies.

This paper is organized as follows. 
In Sec. \ref{sec:model}, the model is presented. In Sec. \ref{sec:constraints}, various constraints imposed on the parameters by several measurements are examined. Sec. \ref{sec:LHC} is devoted to LHC phenomenology of the model and the strategy to test it. In Sec. \ref{sec:sum} we summarize our work.

\section{The model} \label{sec:model}

To accommodate the $Z'$ boson and vectorlike particles, the SM is extended to incorporate a U(1)$_X$ gauge group.
The U(1)$_X$ charges of all the SM fields are assumed to be 0.
Under $\text{SU}(3)_C\times \text{SU}(2)_L\times \text{U}(1)_Y\times \text{U}(1)_X$, the newly introduced fields are in the following representations \cite{Sierra:2015fma},
\begin{equation}
Q_{L,R}=(3,2,1/6,1)~,
\quad
L_{L,R}=(1,2,-1/2,1)~,
\end{equation}
\begin{equation}
\phi=(1,1,0,1)~,
\end{equation}
where $Q=(U, D)$ and $L=(N,E)$ are vectorlike fermions and $\phi$ is a scalar that develops a vacuum expectation value (VEV) $\langle \phi \rangle = \frac{1}{\sqrt{2}}v_\phi$ that breaks U(1)$_X$ and gives mass to the $Z'$ boson, $m_{Z'}=g_X v_\phi$.
The mixings of $Q_R(L_R)$ with its right-handed SM counterparts are roughly proportional to the SM Yukawa couplings and therefore highly suppressed except for the top quark.
The mixings of $Q_L(L_L)$ with left-handed SM quarks(leptons) can be large, since they are charged under the SM gauge group in the same way.
As a consequence, the gauge interactions of SM fermions are almost intact and electroweak precision measurements put few constraints on the mixing angles.
We can write down Dirac mass terms for the vectorlike fermions and their Yukawa couplings to SM counterparts,
\begin{equation}
\mathcal L \supset -m_Q \bar Q_L Q_R-m_L \bar L_L L_R-\lambda_Q^i \bar Q_R \phi q^i - \lambda_L^i \bar L_R \phi \ell^i+\text{h.c.}~,
\end{equation}
where $\ell^i$ and $q^i$ represent the left-handed SM lepton and quark doublets, respectively. $\lambda_L^{1,2,3}$ and $\lambda_Q^{1,2,3}$ are denoted by $\lambda_L^{e,\mu,\tau}$ and $\lambda_Q^{d,s,b}$ below.
Since the terms involving $\lambda_L^\mu$ and $\lambda_Q^{b,s}$ are sufficient to provide the couplings that lead to $b \to s$ anomalies, $\lambda_L^{e,\tau}$ and $\lambda_Q^{d}$ are set to be 0. This eliminates all lepton flavor violating (LFV) processes mediated by the $Z'$ boson and simplifies the model substantially yet still gives rise to rich phenomenology. 

After the new scalar field acquires a VEV, the mass matrices for charged leptons and down-type quarks can be written as
\begin{equation}
\mathcal{M}_E = \bordermatrix{
~ & E_L & \mu_L \cr
E_R & m_L & \frac{\lambda_L^\mu v_\phi}{\sqrt{2}} \cr
\mu_R & 0 & \frac{y_\mu v}{\sqrt{2}}
}~,
\quad \mathcal{M}_D = \bordermatrix{
~ & D_L & s_L & b_L \cr
D_R & m_Q & \frac{\lambda_Q^s v_\phi}{\sqrt{2}} & \frac{\lambda_Q^b v_\phi}{\sqrt{2}} \cr
s_R & 0  & \frac{y_s v}{\sqrt{2}} & 0 \cr
b_R & 0 & 0 & \frac{y_b v}{\sqrt{2}} \cr
}~,
\end{equation}
where $y_{\mu, s,b}$ are SM Yukawa couplings and $v$ is the Higgs VEV. The up-type quark mass matrix is similar to $\mathcal{M}_D$. In the limit of $m_Q \gg y_{\mu, s,b}~v$, the masses for the new lepton and quark are 
\begin{equation}	
m_E=\sqrt{m_L^2+\frac{|\lambda_L^\mu|^2 v_\phi^2}{2}}~,
\quad
m_D=\sqrt{m_Q^2+\frac{\left(|\lambda_Q^b|^2+|\lambda_Q^s|^2\right)v_\phi^2}{2}}~.
\end{equation}
The effective couplings of the $Z'$ boson to SM fermions are induced by mixings between vectorlike and SM fermions,
\begin{equation}
\mathcal L\supset g_{f_i f_j} Z'_\mu \bar f_i\gamma^\mu P_L f_j~,
\end{equation}
with 
\begin{equation}
g_{bs}=\frac{\lambda_Q^b\lambda_Q^{s*} v_\phi^2}{2m_D^2}g_X~,
\quad
g_{bb}=\frac{|\lambda_Q^b|^2 v_\phi^2}{2m_D^2}g_X~,
\quad
g_{ss}=\frac{|\lambda_Q^s|^2 v_\phi^2}{2m_D^2}g_X~,
\quad
g_{\mu\mu}=\frac{|\lambda_L^\mu|^2 v_\phi^2}{2m_E^2}g_X~.  \label{eq:cps}
\end{equation}
The remaining nonzero couplings $g_{\nu_\mu\nu_\mu}$, $g_{cc}$, $g_{tt}$ and $g_{tc}(g_{ct})$ are not independent: $g_{\nu_\mu\nu_\mu}=g_{\mu\mu}$, $g_{cc}=g_{ss}$ with $m_c$ being neglected, and $g_{tt}=g_{bb}$, $g_{tc}=g_{bs}$ provided that $m_Q\gg m_t$. Note that $g_{bb}$, $g_{ss}$ and $g_{bs}$ are correlated, there is always $g_{bb}g_{ss}=|g_{bs}|^2$.

\section{Low energy constraints} \label{sec:constraints}

The measurements of $R_{K^{(*)}}$ supplemented by a few other low energy measurements mentioned below imply a correlation between the mass of $Z'$ and its couplings to the SM fermions. 

The effective Hamiltonian describing $b\to s\ell\ell$ transitions is conventionally written as \cite{Altmannshofer:2017yso}
\begin{equation}
\mathcal H_\text{eff}=-\frac{4G_F}{\sqrt{2}}V_{tb}V_{ts}^*\frac{e^2}{16\pi^2}\sum_{i,\ell}(C_i^\ell \mathcal O_i^\ell+C_i^{\prime\ell} \mathcal O_i^{\prime\ell})+\text{h.c.}~,
\end{equation}
with the following four-fermion interactions:
\begin{equation}
\mathcal O_9^\ell=(\bar s\gamma_\mu P_L b)(\bar \ell \gamma^\mu \ell)~,
\quad
\mathcal O_{10}^\ell=(\bar s\gamma_\mu P_L b)(\bar \ell \gamma^\mu\gamma_5 \ell)~,
\end{equation}
\begin{equation}
\mathcal O_9^{\prime\ell}=(\bar s\gamma_\mu P_R b)(\bar \ell \gamma^\mu \ell)~,
\quad
\mathcal O_{10}^{\prime\ell}=(\bar s\gamma_\mu P_R b)(\bar \ell \gamma^\mu\gamma_5 \ell)~.
\end{equation}
The primed Wilson coefficients do not receive significant SM contributions while the unprimed can be split into two parts, the SM and the NP ones,
\begin{align}
C^{\ell}_9 = C^{\ell,\text{SM}}_9 + C^{\ell, \text{NP}}_9~, \quad C^{\ell}_{10} = C^{\ell,\text{SM}}_{10} + C^{\ell, \text{NP}}_{10}~. 
\end{align}
Many groups have performed global fits to the data on $b \to s\ell\ell$ transitions \cite{Alok:2017jgr,Capdevila:2017bsm,Altmannshofer:2017yso,DAmico:2017mtc,Geng:2017svp,Ciuchini:2017mik,Alok:2017sui}.
One of the favored scenarios can be exactly implemented in our model, where new particles only couple to left-handed quarks and left-handed muons, i.e., $C_9^{\mu,\text{NP}}=-C_{10}^{\mu,\text{NP}}$ while all other coefficients of NP remain zero. To be specific, the effective Hamiltonian for anomalous $b\to s\ell\ell$ transitions is
\begin{equation}
\mathcal H_\text{eff}^\text{NP}=-\frac{g_{bs}g_{\mu\mu}}{m_{Z'}^2}(\bar s\gamma_\mu P_L b)(\bar \ell \gamma^\mu P_L \ell)+\text{h.c.}~.
\end{equation}
The best fit point that takes into account only LFU observables instead of all available data requires $C_9^{\mu,\text{NP}}=-C_{10}^{\mu,\text{NP}}=-0.63$ \cite{Altmannshofer:2017yso}, which translates into 
\begin{equation}
\frac{m_{Z'}^2}{g_{bs}g_{\mu\mu}} \simeq 947~\text{TeV}^2~. \label{eq:bestfit}
\end{equation}

There are two relevant constraints in the parameter space around the best fit point, one from $B_s-\bar B_s$ mixing \cite{Altmannshofer:2014rta} and the other from the neutrino trident production \cite{Altmannshofer:2014pba}. The former puts a bound on $m_{Z'}$ over $g_{bs}$ while the latter pertains to $m_{Z'}$ over $g_{\mu\mu}$,
\begin{equation}
\frac{m_{Z'}}{g_{bs}} \gtrsim 244~\text{TeV}~,
\quad
\frac{m_{Z'}}{g_{\mu\mu}} \gtrsim 0.47~\text{TeV}~. \label{eq:bsbs}
\end{equation}
Combining these constraints with the relations in Eq. (\ref{eq:cps}) gives 
\begin{equation}
v_\phi = \frac{m_{Z'}}{g_X}<\frac{m_{Z'}}{g_{\mu\mu}} = \frac{2m_E^2}{|\lambda_L^\mu|^2 v_\phi} \lesssim \frac{947}{244}~\text{TeV}~. \label{eq:mzOg}
\end{equation}
Perturbativity requires the NP couplings to be less than $\sqrt{4\pi}$. Moreover, if we require $g_X$ not to hit a Landau pole below the Planck scale, utilizing the one-loop beta function for $g_X$, we obtain
\begin{equation}
g_X(m_{Z'})<\left(\frac{11}{8\pi^2}\log\frac{M_\text{Pl}}{m_{Z'}}\right)^{-\frac{1}{2}}\simeq 0.45~.
\end{equation}
The constraints on $g_X$ and $\lambda_L^\mu$ then indicate that the masses of the $Z'$ boson and the vectorlike lepton are bounded from above \footnote{Substituting $m_{Z'}=g_X v_\phi$, $g_{bs}<\frac{2\pi v_\phi^2}{m_D^2}g_X$ and $g_{\mu\mu}<g_X$ into Eq. (\ref{eq:bestfit}), we obtain another bound: $m_D\lesssim$ 77 TeV.}
\begin{equation}
m_{Z'} \lesssim 3.9g_X~\text{TeV}\lesssim 1.8~\text{TeV}~,
\end{equation}
\begin{equation}
m_E \lesssim 2.7|\lambda_L^\mu|~\text{TeV}\lesssim 9.6~\text{TeV}~. 
\end{equation}
Note that after mixing with the up-type vectorlike quark $U$, the top quark mass is slightly smaller than $\frac{1}{\sqrt{2}}y_t v$, but still lies within the uncertainty of current measurements.

\section{LHC phenomenology}\label{sec:LHC}

\begin{figure}[htbp]
\centering
\includegraphics[width=0.5\textwidth]{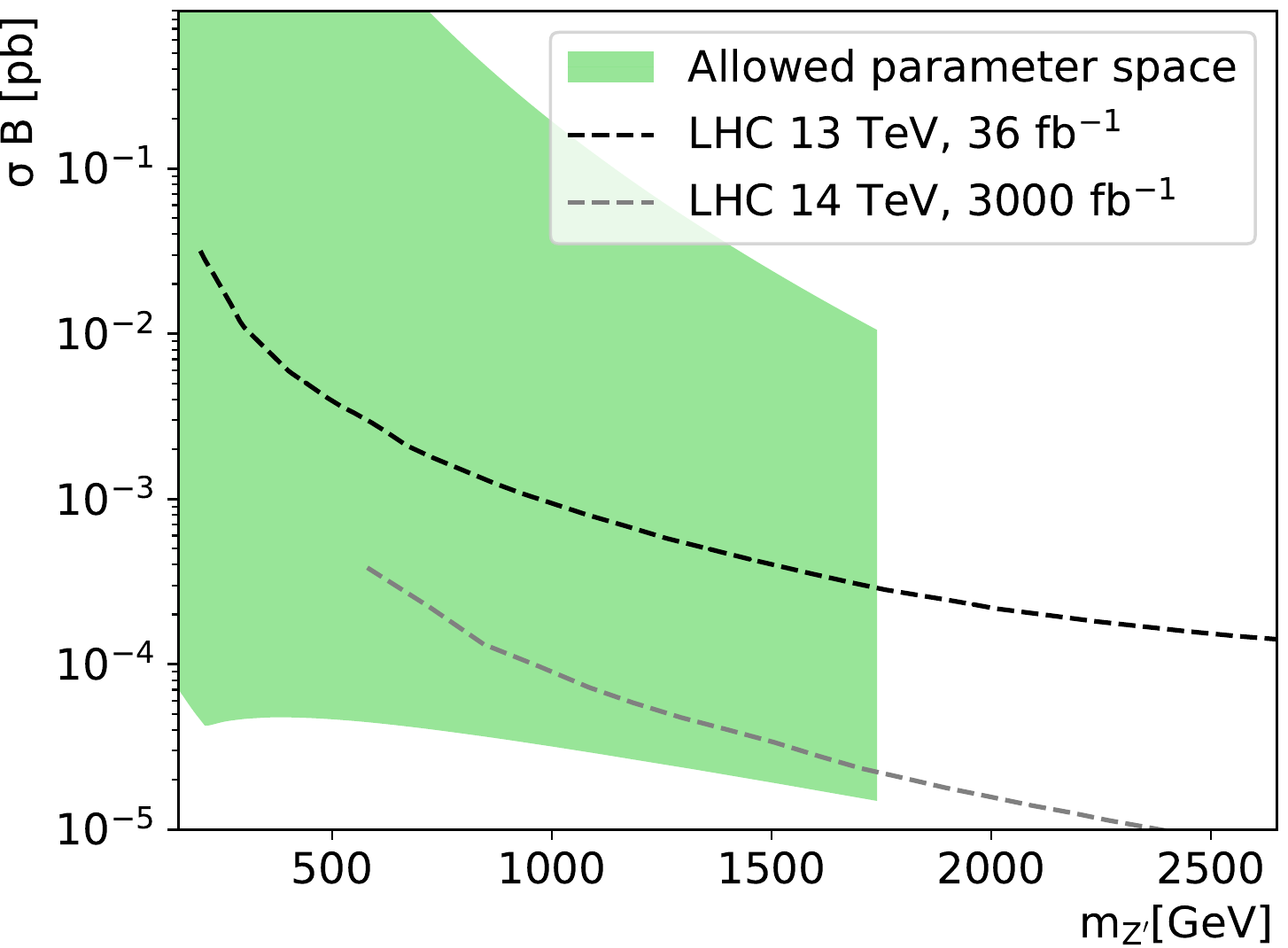}
\caption{\label{fig:sigma}The allowed range of $\sigma(pp\to Z')\times \text{BR}(Z'\to \mu^+\mu^-)$ with respect to the mass of $Z'$ is shown by the green band. The expected exclusion limits at 95\% C.L. by the CMS collaboration using the LHC data at 13 TeV with 36 fb$^{-1}$ are shown by the dashed black curve. And the extrapolated exclusion limits at the 14 TeV LHC with 3000 fb$^{-1}$ are shown by the dashed gray curve.}
\end{figure}

The aforementioned constraints can be used to restrict $\sigma(pp\to Z')\times \text{BR}(Z'\to \mu^+\mu^-)$ which is the true observable concerning a direct collider search.
Larger $g_{f_i f_j}$ with $f_i=c,s,t,b$ typically mean larger $\sigma(pp\to Z')$ but lower BR($Z'\to \mu^+\mu^-$) and vice versa. Each of the processes $cc\to Z'$, $ss\to Z'$, $bb\to Z'$ and $bs\to Z'$ contributes a certain fraction of $\sigma(pp\to Z')$, and $\sigma\times$BR reaches the lower bound approximately when $g_{f_i f_j}$ are adjusted accordingly to minimize $\sigma(pp\to Z')$ while BR$(Z'\to \mu^+\mu^-) \simeq 50\%$, because couplings of $Z'$ to quarks are negligible compared to $g_{\mu\mu}$ at this point. The upper bound does not necessarily matter because that region of parameter space has already been ruled out by current searches. The boundaries of parameter space consistent with the constraints discussed in Sec. \ref{sec:constraints} are presented in the $\sigma\times$BR vs $m_{Z'}$ plane in Fig. \ref{fig:sigma}.

\begin{figure}[htbp]
\includegraphics[width=0.5\textwidth]{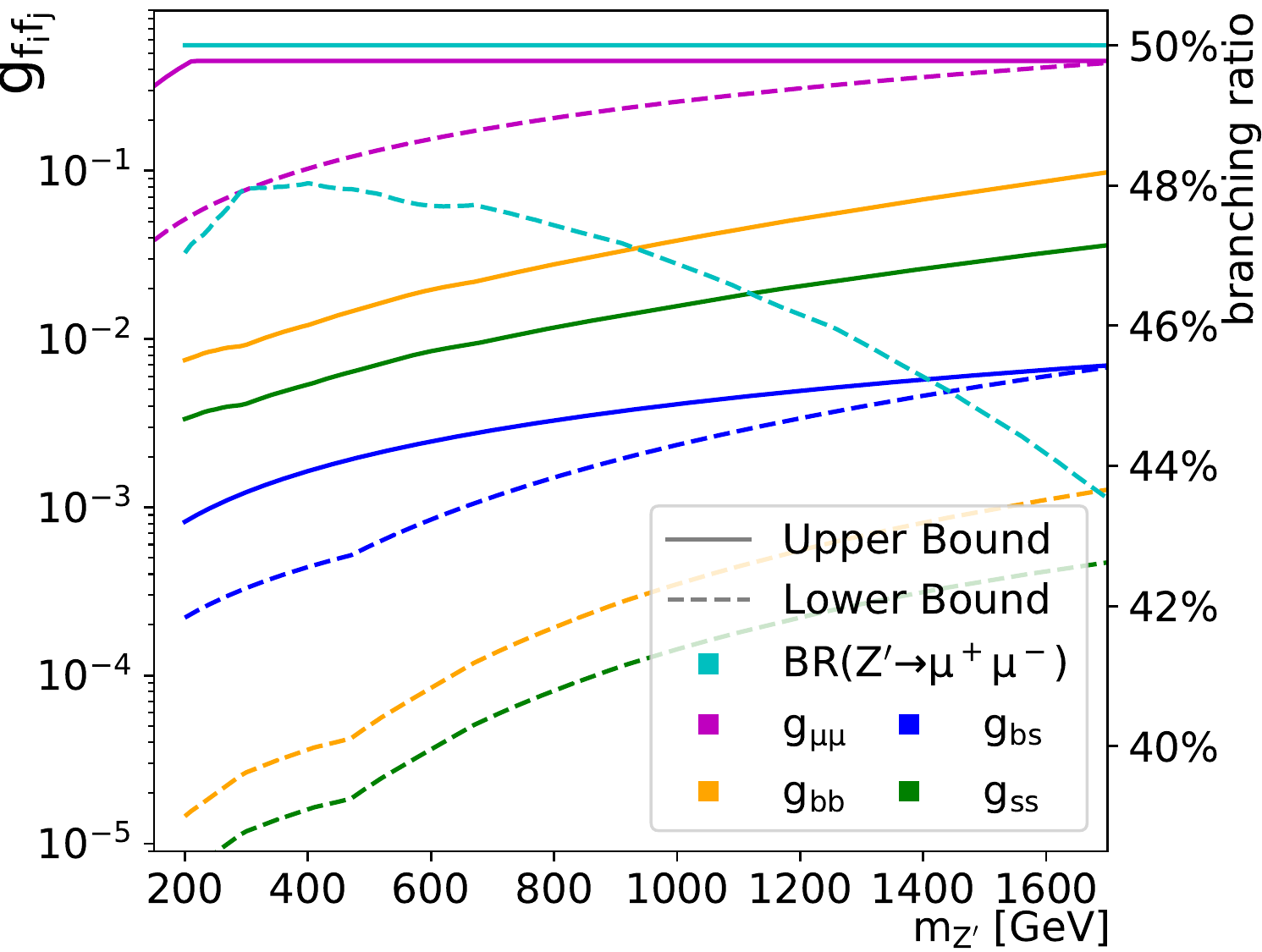}
\caption{\label{fig:gff}The bounds on the couplings of $Z'$ to the SM fermions (left $y$ axis) and the branching ratio for $Z'\to\mu^+\mu^-$ (right $y$ axis), taking into account the exclusion limits on $Z'$ in the dimuon channel at the LHC. Magenta, orange, green, blue, and cyan correspond to $g_{\mu\mu}$, $g_{bb}$, $g_{ss}$, $g_{bs}$, and BR$(Z'\to\mu^+\mu^-)$ respectively. The solid and dashed lines represent upper and lower bounds respectively. It should be noted that $g_{bb}$, $g_{ss}$, and $g_{bs}$ are not independent; in other words, they cannot reach the upper or lower bounds at the same time.}
\end{figure}

The search for $Z'$ in the dimuon final state has been performed by both the ATLAS \cite{Aaboud:2017buh} and CMS \cite{Sirunyan:2018exx} collaborations at the LHC. The expected exclusion limits by CMS using 36 fb$^{-1}$ of data collected at 13 TeV are shown by the dashed black curve in Fig. \ref{fig:sigma}. The region above the curve covering half of the viable parameter space is excluded. Further constraints on $g_{f_if_j}$ can be derived from the exclusion limits as follows. Combining Eqs. (\ref{eq:bestfit}) and (\ref{eq:bsbs}), we obtain both the upper and lower bounds on $g_{\mu\mu}$ and $g_{bs}$. With $g_{\mu\mu}$ saturating the lower bounds, larger $g_{bb}$ or $g_{ss}$ would cause $\sigma\times$BR to exceed the exclusion limits, hence the upper bounds on $g_{bb}$ and $g_{ss}$. Considering that $g_{bb}$, $g_{ss}$ and $g_{bs}$ are correlated, i.e., $g_{bb}g_{ss}=g_{bs}^2$, the lower bounds on $g_{bb}$ and $g_{ss}$ are straightforward. The results are collected in Fig. \ref{fig:gff}. These couplings are directly related to the branching ratio for $Z'\to\mu^+\mu^-$ \footnote{The exact expression in consideration of the top quark mass is slightly modified, though hardly affects the numerical results.}
\begin{equation}
\text{BR}(Z'\to\mu^+\mu^-)\simeq \frac{g_{\mu\mu}^2}{2g_{\mu\mu}^2+6g_{ss}^2+6g_{bs}^2+6g_{bb}^2}~,
\label{eq:zpbrmm}
\end{equation}
whose upper and lower bounds are plotted in Fig. \ref{fig:gff} as well. Because $g_{\mu\mu}\gg g_{bs}$ and $g_{bb}g_{ss}=g_{bs}^2$, the 50\% upper bound is trivial and is reached when $g_{ss,bb,bs}$ are of the same order.
$Z'$ decays dominantly to leptons if $g^2_{\mu\mu} \gg 3\max\left(g^2_{bb},~g^2_{ss}\right)$. In fact, the leptonic branching ratio is at least 90\% for $m_{Z'}$ under 1.5 TeV.

To show the prospect for a discovery in the dimuon channel, we extrapolate the exclusion limits to the 14 TeV LHC with an integrated luminosity of 3000 fb$^{-1}$ using a method dedicated to resonance searches \cite{Thamm:2015zwa,Allanach:2017bta}. The result merely serves as a rough estimate of future collider sensitivity; we expect the agreement between the extrapolation and a cut-and-count analysis within a factor of 2 according to Ref. \cite{Thamm:2015zwa}. The sensitivity to $\sigma\times$BR can be improved by 1 order of magnitude, but there is still plenty of parameter space left, even a very light $Z'$ with a few hundred GeV may possibly escape the resonance search. So it demands other strategies to test this model, and we show that the search for the new charged lepton $E^\pm$ provides a complementary probe.

Searches for vectorlike leptons have been studied within various theoretical frameworks \cite{Gross:2010ce,Dermisek:2014qca,Kumar:2015tna}.
In our model, the interactions between $E^\pm$ and SM gauge bosons are given by
\begin{equation}
\mathcal L \supset g\frac{\lambda_L^\mu v_\phi}{4m_E}\left(\frac{y_\mu v}{\cos\theta_W m_E} Z_\mu \bar\mu_R \gamma^\mu E_R + \frac{(y_\mu v)^2 m_L}{m_E^3} W^+_\mu \bar v_\mu \gamma^\mu E_L\right)+O(y_\mu^3)+\text{h.c.}~,
\end{equation}
where $y_\mu$ is the muon Yukawa coupling and $v$ the SM Higgs VEV. As a result, the $E^\pm \to \mu^\pm Z$ ($\propto y_\mu$) and $E^\pm \to \nu W^\pm$ ($\propto y_\mu^2$) decay channels are highly suppressed. 
Moreover, the $E^\pm\to N W^\pm$ decay channel is kinematically impossible as the mass difference between $E$ and $N$ is of order $m_\mu$,
\begin{equation}
\Delta m_L=\frac{\lambda_L^\mu v_\phi y_\mu v}{2m_E}=\frac{\lambda_L^\mu v_\phi}{\sqrt{2}m_L}m_\mu~.
\end{equation}

There are only two major decay channels left,\footnote{Because to the extra scalar boson $\phi$ in our model, there is another decay mode $E^\pm\to \mu^\pm \phi$ if kinematically allowed. The reason why we do not consider it in this work is that the mass of $\phi$ is a free parameter in the model; a light scalar field $\phi$ that mixes with the SM Higgs is stringently constrained; it can only give rise to signatures similar to those of the $E^\pm\to \mu^\pm H$ channel.} $E^\pm\to \mu^\pm H$ and $E^\pm\to \mu^\pm Z^{\prime (*)}$ with either an on-shell or an off-shell $Z^{\prime}$, which subsequently decays into a pair of muons at least 45\% of the time for $m_{Z'}<1.5$ TeV as governed by Eq. (\ref{eq:zpbrmm}).
Through the $E^\pm\to \mu^\pm Z^{\prime (*)}(\rightarrow \mu^+\mu^-)$ channel, 6-muon final states can be produced at the collider. In contrast to the production of a single $Z'$, which are bounded by very small NP couplings, $g_{ss}$, $g_{bs}$, and $g_{bb}$, the $E^\pm$ pair production is practically governed by SM gauge couplings the same as those of $e/\mu$ to $\gamma$ and $Z$. More importantly, the 6-muon signature is almost free from the SM background at hadron colliders \cite{Kumar:2015tna}. The region of parameter space that predicts $\mathcal{O}(10)$ events can be probed with a high significance.
The total number of 6-muon events at a hadron collider is determined by $m_E$, $m_{Z'}$, $g_{f_if_j}$ (mostly $g_{\mu\mu}$ since the others are negligible in the considered region), and the mixing angle between the Higgs boson and $\phi$, $\theta_{H-\phi}$, which diminishes BR$(E^\pm\to \mu^\pm Z^{\prime (*)})$. 
On top of the boundaries depicted in Fig. \ref{fig:gff}, $g_{\mu\mu}$ is subject to the bound,
\begin{equation}
g_{\mu\mu}=\frac{|\lambda_L^\mu| m_{Z'}\sqrt{m_E^2-m_L^2}}{\sqrt{2}m_E^2}<\frac{\sqrt{2\pi}~m_{Z'}}{m_E}~.\label{eq:guu}
\end{equation}

\begin{figure}[htbp]
\begin{subfigure}[t]{\textwidth}
\includegraphics[width=0.328\textwidth]{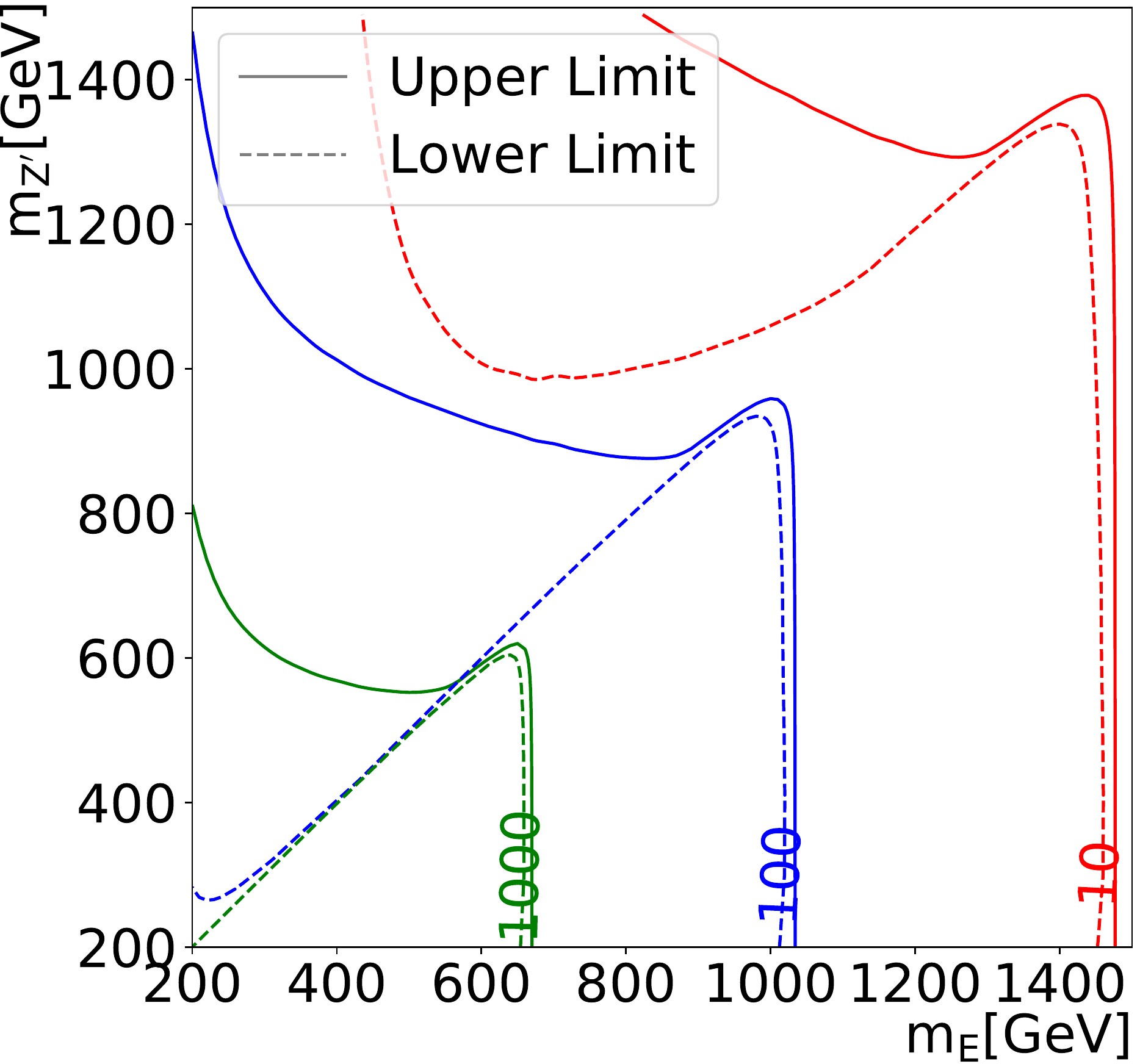}
\hfill
\includegraphics[width=0.328\textwidth]{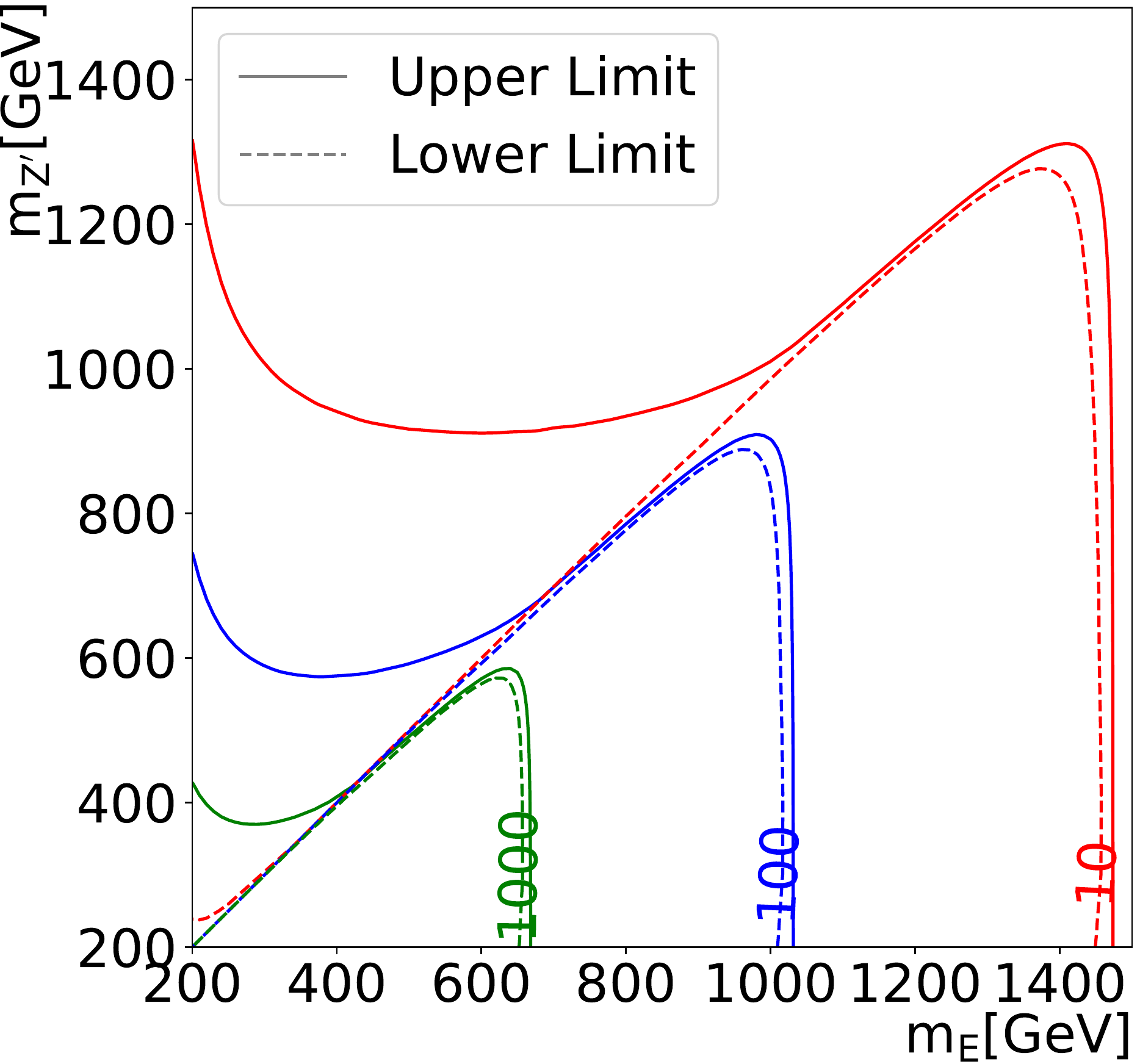}
\hfill
\includegraphics[width=0.328\textwidth]{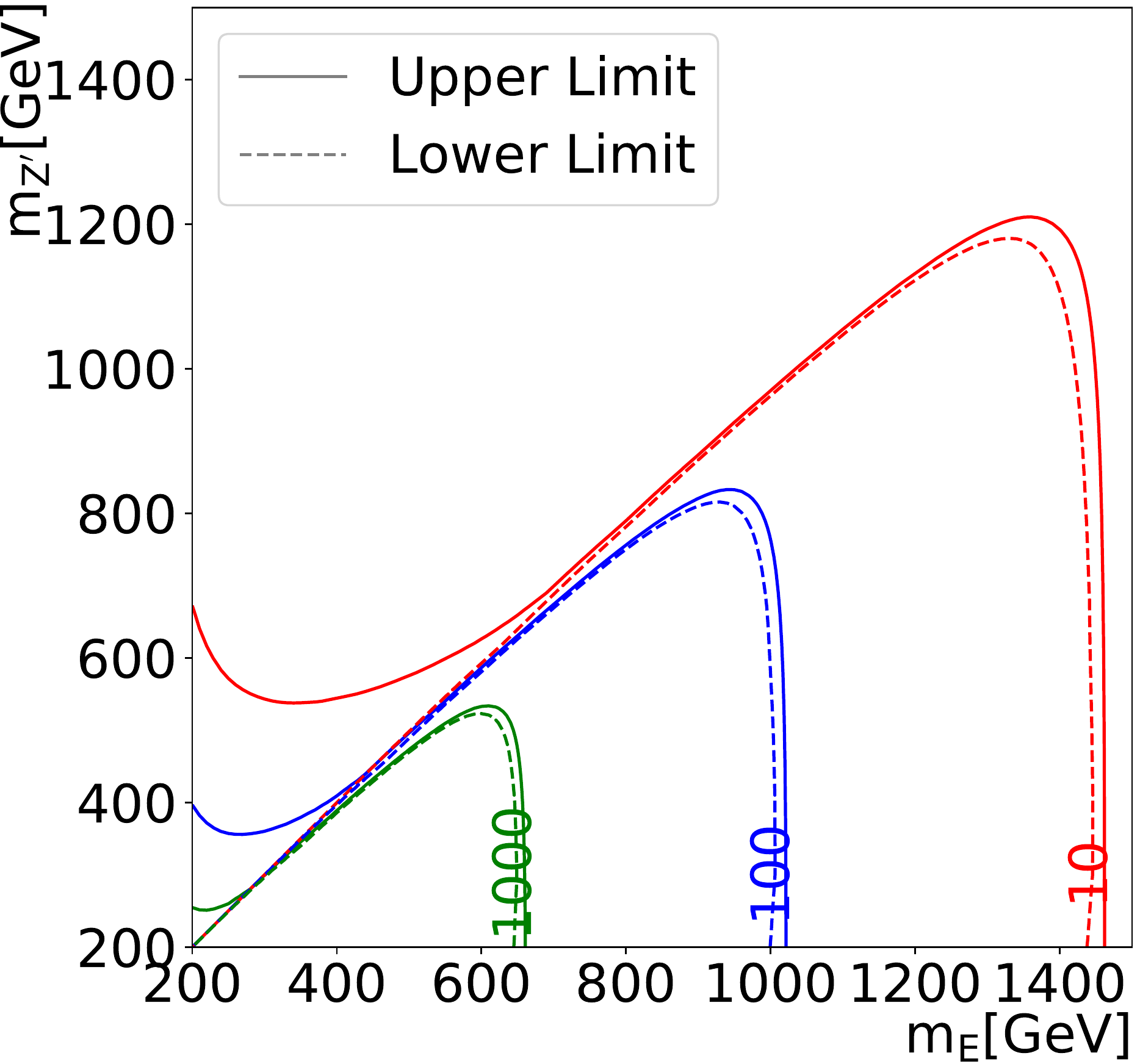}
\end{subfigure}
\par\bigskip
\begin{subfigure}[t]{\textwidth}
\includegraphics[width=0.328\textwidth]{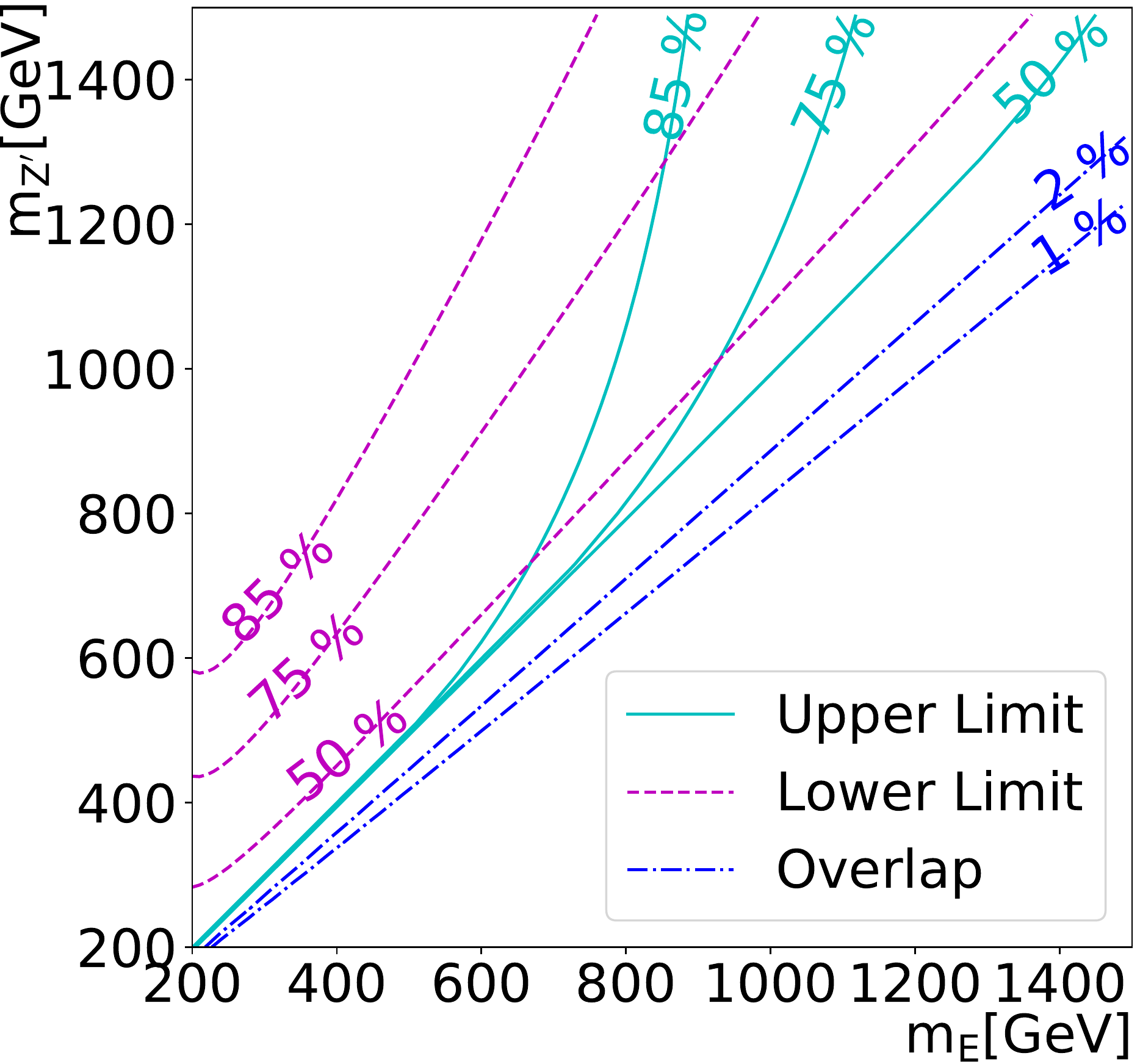}
\hfill
\includegraphics[width=0.328\textwidth]{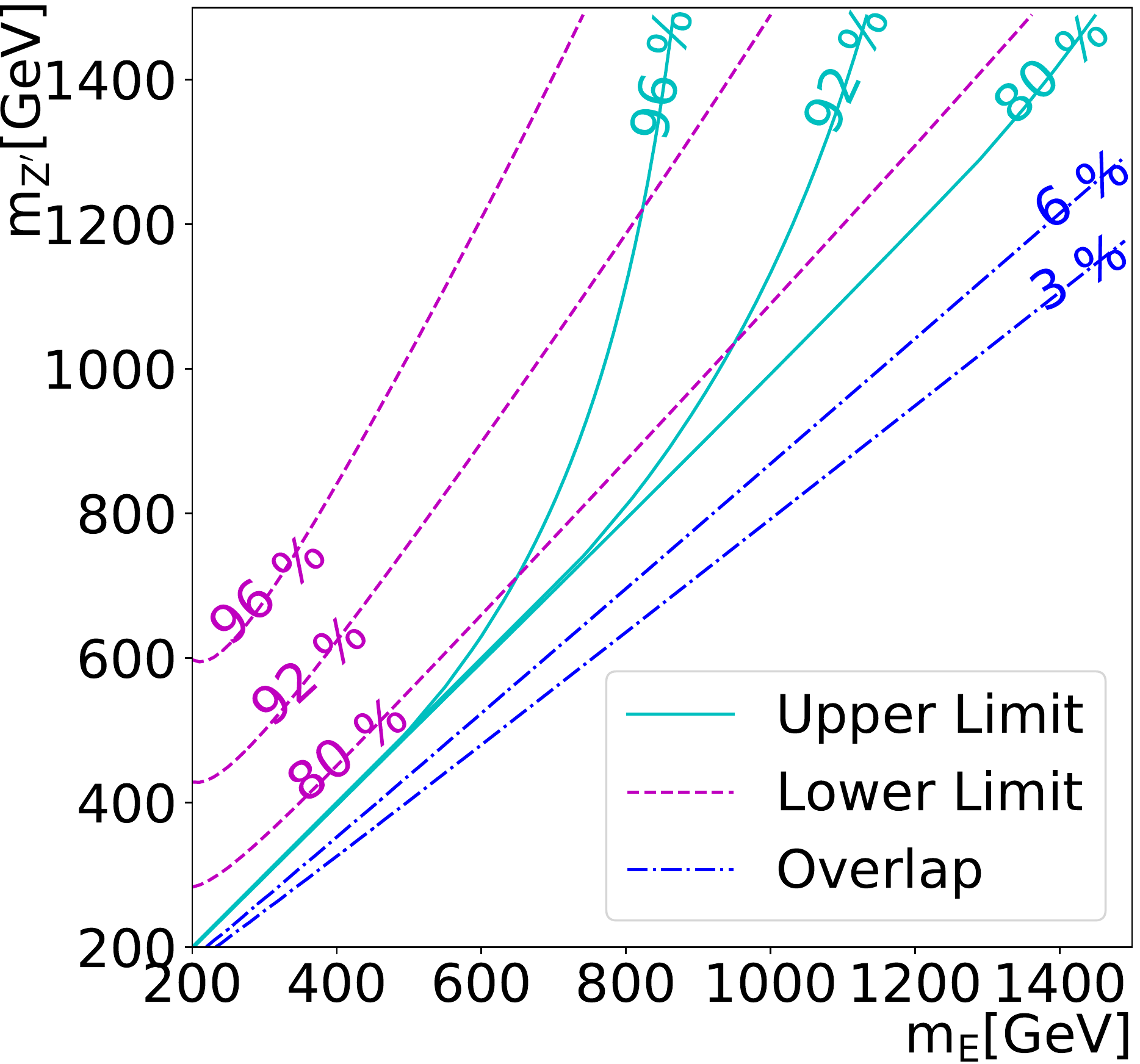}
\hfill
\includegraphics[width=0.328\textwidth]{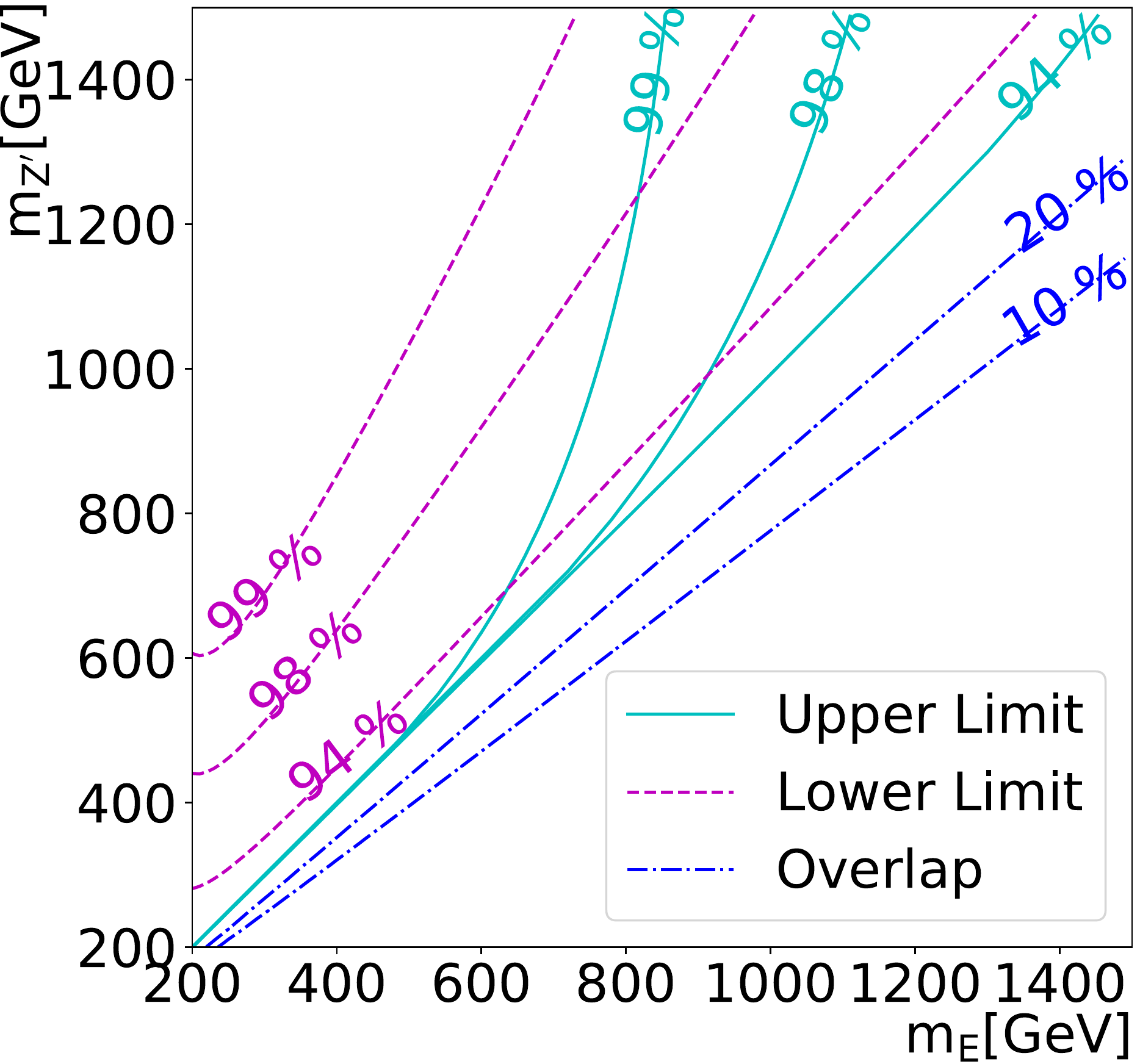}
\end{subfigure}
\caption{\label{fig:br}Top: The expected number of 6-muon events at the LHC at 14 TeV with 3000 fb$^{-1}$, where the contours of 1000, 100 and 10 events are plotted. Only decays into SM fermion pairs are taken into account in the calculation of $\Gamma_{Z'}$. Bottom: The contours of BR$(E^\pm\to \mu^\pm H)$. The mixing angle between the Higgs boson and $\phi$ is set to $\sin \theta_{H-\phi}$ = 0.05 (left), 0.1 (middle), 0.2 (right), respectively. In all the plots, the upper limits are represented by solid lines and lower limits by dashed lines. The upper and lower limits on BR$(E^\pm\to \mu^\pm H)$ almost coincide in the region where $m_{Z'}<m_E$, and thus are represented by dash-dotted lines instead. Owing to the fact that all $g_{f_if_j}$ except $g_{\mu\mu}$ are negligible in the displayed region, basically the number of 6-muon events reaches its upper limits while BR$(E^\pm\to \mu^\pm H)$ reaches its lower limits and vice versa.}
\end{figure}

The upper (solid contours) and lower (dashed contours) limits on the expected number of 6-muon events at the LHC at a center-of-mass energy of 14 TeV with an integrated luminosity of 3000 fb$^{-1}$ are plotted in the top panels of Fig. \ref{fig:br}, where $\textsc{MadGraph5\_aMC@NLO 2.6.0}$ \cite{Alwall:2014hca} has been used to evaluate the production cross section of the $E^\pm$ pair.
Needless to say, both the upper and lower limits are highly dependent on $m_E$.
One remarkable feature of these plots is that the number of events hardly changes with $m_{Z'}$ as long as $Z'$ is on shell but drops abruptly when $Z'$ goes off shell, especially in the lower limits.

\begin{figure}[htbp]
\centering
\includegraphics[width=0.45\textwidth]{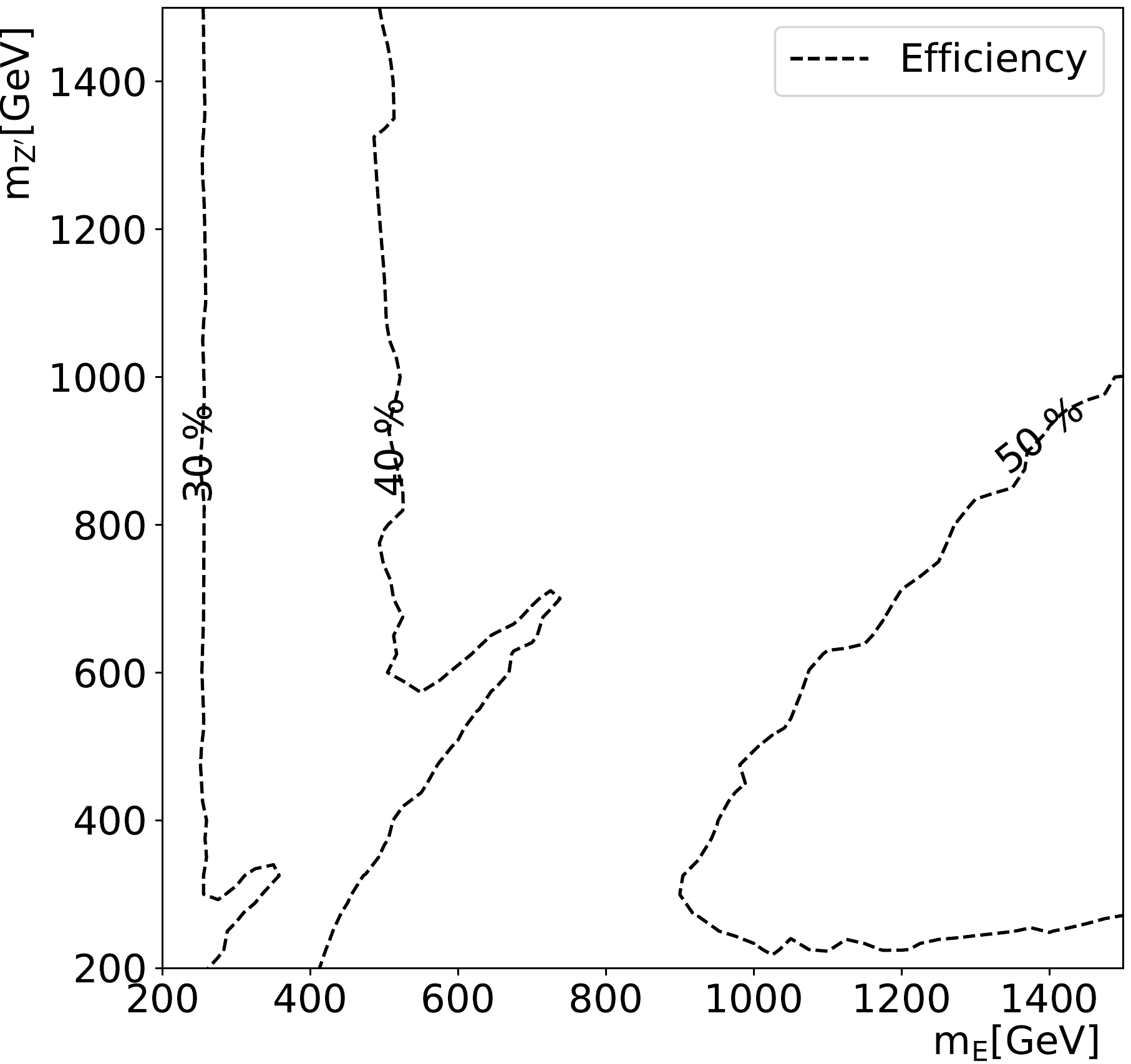}
\caption{\label{fig:effi} 6-muon events selection efficiency.}
\end{figure}

The plots in Fig. \ref{fig:br} do not take detector or parton showering effects into consideration. For leptonic final states, showering effects are negligible.
We estimate the efficiency for the CMS detector to identify a 6-muon final state as follows: each muon identification efficiency is 95\% for $p_T>10$ GeV, $|\eta|<2.4$, and $\Delta R(\mu,\mu)>0.4$, and 0 otherwise. The overall selection efficiency is shown in Fig. \ref{fig:effi}, where the decay width of $Z'$ is fixed to $\Gamma_{Z'}/m_{Z'}=1\%$. Typically around 40\%-50\% of the total events will be selected.
In view of the top panels of Fig. \ref{fig:br} and the extremely low SM background \cite{Kumar:2015tna}, we expect that the charged vectorlike lepton as heavy as 1400 GeV can be probed in the parameter space where $m_{Z'}<m_{E}$.

\begin{figure}[htbp]
\centering
\includegraphics[width=0.5\textwidth]{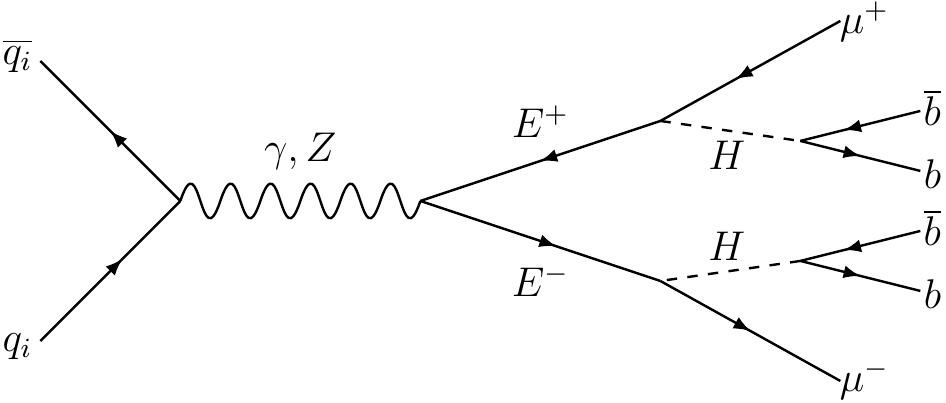}
\caption{\label{fig:feyndiag} The Feynman diagram for the production and decays of a pair of vectorlike leptons.}
\end{figure}

The upper and lower limits of BR$(E^\pm \to \mu^\pm H)$ are plotted in the bottom panels of Fig. \ref{fig:br}. The two limits almost coincide with each other in the region where $m_{Z'}<m_{E}$ and $E^\pm\to \mu^\pm Z'$ is dominant. The branching ratio increases with increasing $m_{Z'}/m_E$ and $\sin \theta_{H-\phi}$. The current Higgs precision measurements still allow $\left|\sin \theta_{H-\phi}\right|\lesssim 0.3$ \cite{Dupuis:2016fda}. With a sizeable scalar mixing ($\sin\theta_{H-\phi}> 0.1$) and an off-shell $Z'$, i.e., $m_{Z'}/m_{E}>1$, the $E^\pm \to \mu^\pm H$ decay mode dominates over the $E^\pm\to \mu^\pm Z^{\prime*}$ mode. Even with $\sin \theta_{H-\phi}=0.05$ and $m_{Z'}/m_E=1$, BR$(E^\pm \to \mu^\pm H)$ could still be as large as 50\%. In the following subsections, we study its collider phenomenology in detail. The signal to be considered is illustrated in Fig. \ref{fig:feyndiag}. The contribution from $s$-channel $Z'$ exchange is ignored, which is justified by the fact that only the second and third generation quarks couple to $Z'$ and their couplings are negligible compared with those to $\gamma$ and $Z$ except for a very heavy $Z'$.

We have used $g_X<0.45$ and $|\lambda_Q^{s,b}|, |\lambda_L^\mu|< \sqrt{4\pi}$ by perturbativity arguments, though in principle, these constraints could be violated or tightened, which in turn changes some of the results above. Note that the upper bounds on the masses of new particles depend on these parameters linearly, $m^\text{upp}_{Z'}\propto g_X$, $m^\text{upp}_E\propto|\lambda_L^\mu|$, $m^\text{upp}_D\propto\sqrt{|\lambda_Q^s\lambda_Q^b|}$.
Other than that, changing the bounds on $\lambda_Q^{s,b}$ has no impact, while changing the bound on $\lambda_L^\mu$ slightly modifies Fig. \ref{fig:br} through Eq. (\ref{eq:guu}), the differences occur in a restricted region where $m_{Z'}<\frac{\sqrt{2}g_X}{\lambda_L^\mu}m_E$ and are insignificant. Increasing $g_X$ would bring a more dramatic change to Fig. \ref{fig:br}, but only in the region where $m_{Z'}>m_E$ by enhancing $E^\pm\to \mu^\pm Z^{\prime*}$, and potentially reduce the sensitivity to the channel $E^\pm \to \mu^\pm H$.

\subsection{Simulated samples and object reconstruction}

Our signal and background events are generated with $\textsc{MadGraph5\_aMC@NLO 2.6.0}$ \cite{Alwall:2014hca}, in which $\textsc{MadSpin}$ \cite{Artoisenet:2012st} is used for the decays of the vectorlike leptons and SM Higgs boson, and $\textsc{Pythia8}$ \cite{Sjostrand:2014zea} is used to implement parton shower, hadronization and decay of hadrons. The detector effects are simulated by $\textsc{Delphes 3.4.0}$ \cite{deFavereau:2013fsa} with ATLAS configuration card, where the $b$-tagging efficiency has been set to 70\% \cite{ATL-PHYS-PUB-2016-012}, and mistagging rates for the charm- and light-flavor jets are 0.15 and 0.008, respectively. The jet reconstruction is handled by $\textsc{FastJet 3.2.1}$ \cite{Cacciari:2011ma}. 
The signal benchmark points are chosen as $m_E \in [150,~1500]$ GeV with step size of 25 GeV. 
The dominant SM background (BKG) processes for this signal are $t\bar{t}$, $t\bar{t}b\bar{b}$, $t\bar{t}H$ and $t\bar{t}Z$. Their estimated production cross sections at next-to-leading order (NLO) at 14 TeV proton-proton collider \cite{Czakon:2013goa,Kardos:2013vxa,Kardos:2011na,Kulesza:2017ukk} are given in Table \ref{tab:bgs}. 

\begin{table}[htbp]
\centering
\begin{tabular}{|c|c|c|c|c|}
\hline
BKG     & $t\bar{t}$ & $t\bar{t}b\bar{b}$ & $t\bar{t}H$ & $t\bar{t}Z$\\ \hline
Cross section(NLO) & 933 pb & 2636 fb & 611 fb & 1121 fb \\ \hline
\end{tabular}
\caption{\label{tab:bgs} The background cross sections at the 14 TeV LHC. } 
\end{table}

In our analysis, the Higgs bosons are reconstructed with two different methods, and each is suitable for a certain phase space. In the first method, all jets in the final state are reconstructed with anti-$k_t$ algorithm \cite{Cacciari:2008gp} with radius parameter $R=0.4$. Among them, we require at least three $b$-tagged jets for Higgs reconstruction. The combination of the three $b$-tagged jets with the fourth jet that minimizes the mass asymmetry
\begin{equation}
       A=\frac{m_{H_1}-m_{H_2}}{m_{H_1}+m_{H_2}}
\end{equation}
defines two Higgs bosons. We denote them by normal Higgs (NOR Higgs) in the following. 
The second method is devoted to tagging more energetic Higgs bosons, which forms a single jet in the detector. In this case, the jets in the final state are reconstructed by the Cambridge-Aachen (CA) algorithm \cite{Dokshitzer:1997in} with cone size parameter $R=1.4$. The CA jets that fulfill the mass-drop tagger \cite{Butterworth:2008iy} as well as contain at least one $b$-tagged subjet are identified as Higgs jets. They are denoted by substructure Higgs (SUB Higgs).
Each of the reconstructed Higgs bosons is then combined with one of the two muons in the final state to form a vectorlike lepton. Same as above, the combination that minimizes the asymmetry
 \begin{equation}
       B=\frac{m_{E_1}-m_{E_2}}{m_{E_1}+m_{E_2}}
 \end{equation}
 is chosen. 
 
Because to the relatively low efficiency of reconstructing the vectorlike leptons, especially when they are light, we find the stransverse mass of the dimuon system \cite{Lester:1999tx}
\begin{align}
m_{T2} (\mu_1, \mu_2) \equiv \min_{\mathbf{p}_{T1} +\mathbf{p}_{T2} =  \sum \mathbf{p}^j_T } [ \max (m_T(\mathbf{p}(\mu_1),\mathbf{p}_{T1} ) ,m_T(\mathbf{p}(\mu_2),\mathbf{p}_{T2} ) )]\label{mt2}
\end{align}
outperforms the invariant mass of the reconstructed vectorlike lepton in signal and background discrimination. 
Here the transverse mass $m^2_T(\mathbf{p}(\mu_i),\mathbf{p}_{Ti} ) = (E(\mu_i) + \sqrt{ \mathbf{p}_{Ti}^2 + m^2_H})^2 - (\mathbf{p}(\mu_i) + \mathbf{p}_{Ti})^2$ with $m_H$=125 GeV and index $j$ in Eq. (\ref{mt2}) runs over all Higgs constituents.\footnote{We have two Higgs reconstruction methods in parallel. Index $j$ corresponds to four anti-$k_t$ jets in the NOR Higgs method and two CA jets in the SUB Higgs method.} 

\begin{figure}[htbp]
\begin{subfigure}[t]{0.48\textwidth}
\caption{Higgs mass}
\includegraphics[width=\textwidth]{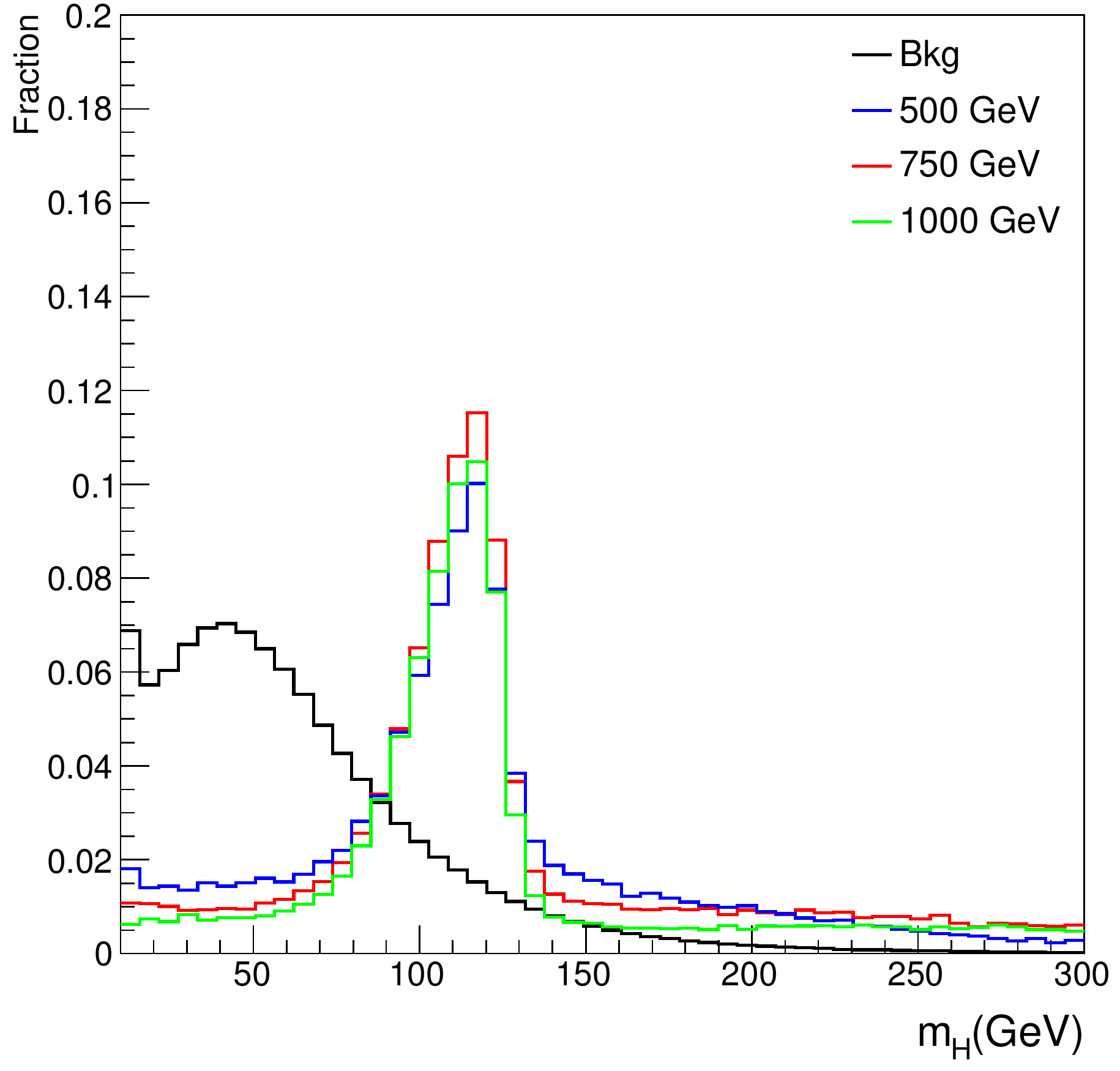}
\end{subfigure}
\hfill
\begin{subfigure}[t]{0.48\textwidth}
\caption{Dimuon stransverse mass}
\includegraphics[width=\textwidth]{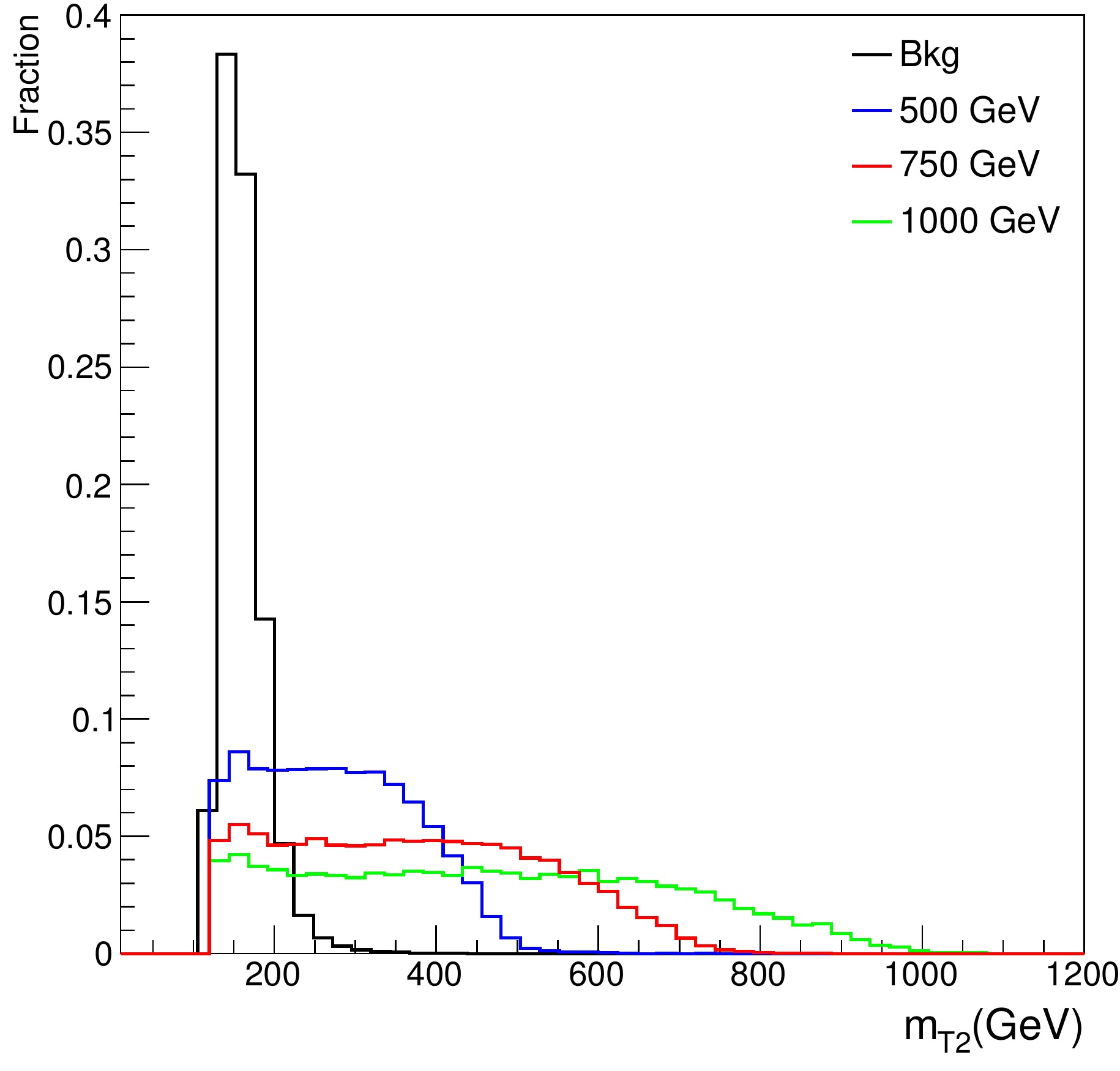}
\end{subfigure}
\par\smallskip
\begin{subfigure}[t]{0.48\textwidth}
\caption{Higgs transverse momentum}
\includegraphics[width=\textwidth]{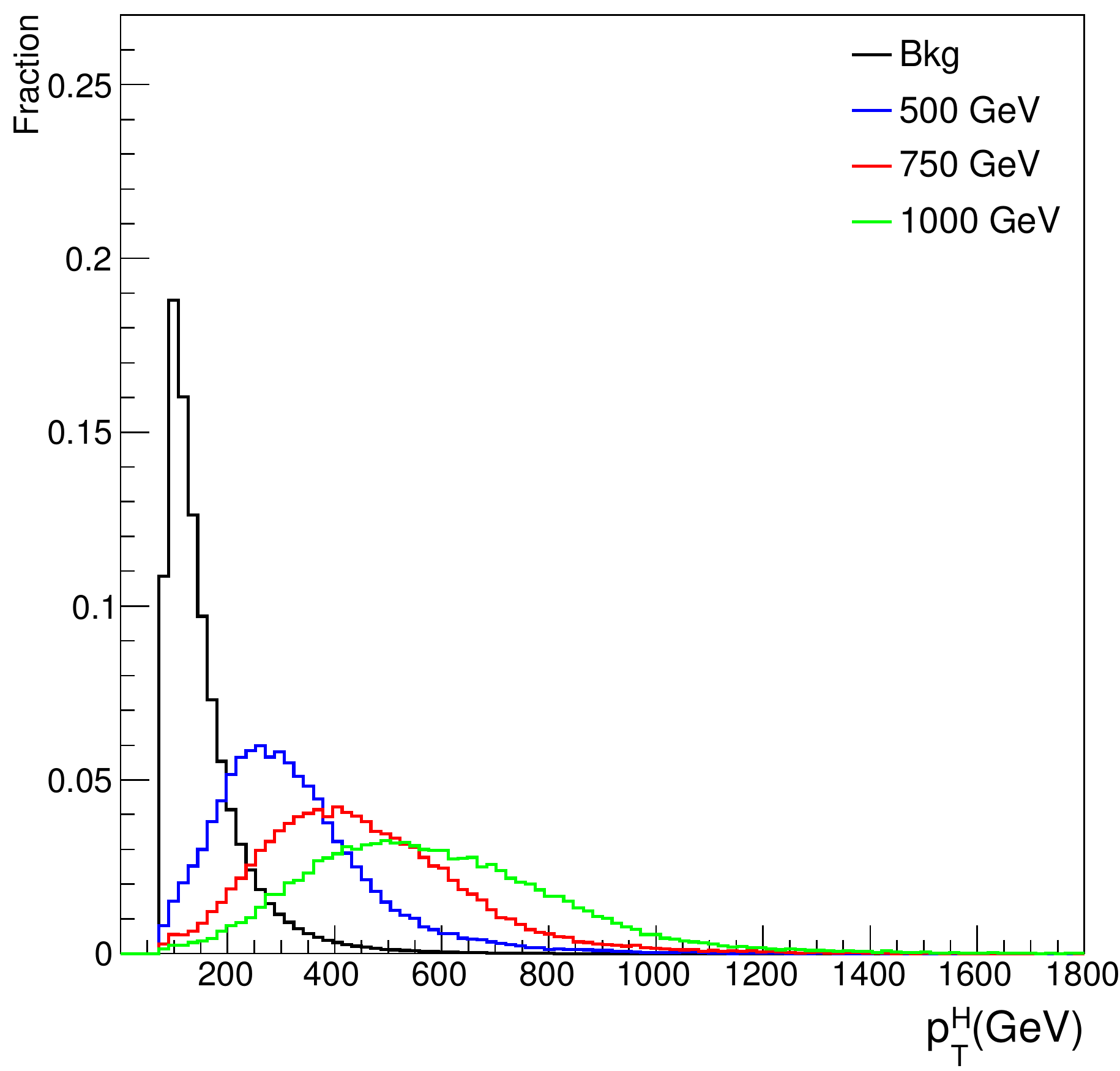}
\end{subfigure}
\hfill
\begin{subfigure}[t]{0.48\textwidth}
\caption{Vectorlike lepton transverse momentum}
\includegraphics[width=\textwidth]{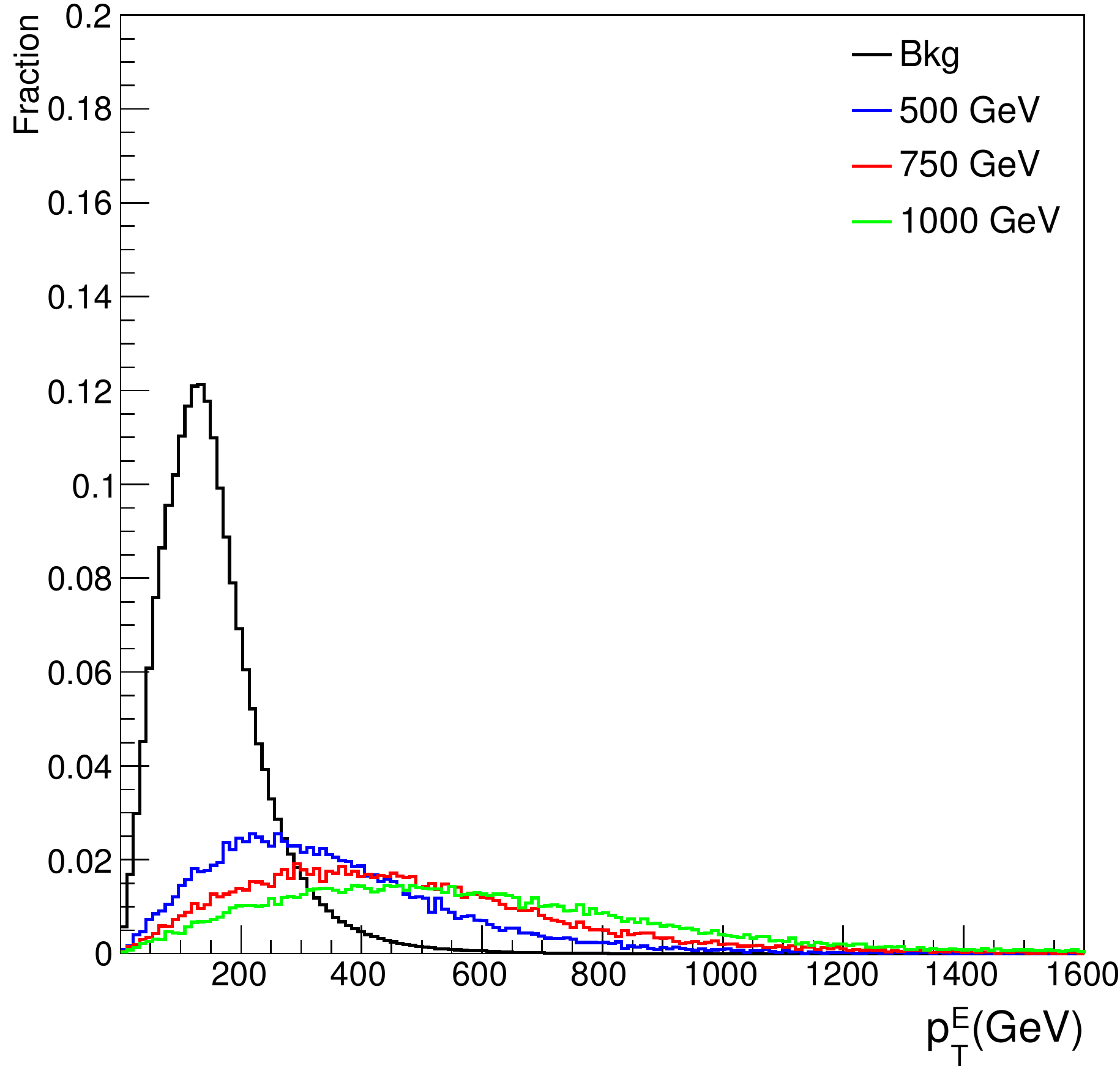}
\end{subfigure}

\caption{\label{fig:dist} Distributions of background events and signal events with $m_{E}$ equal to 500, 750, and 1000 GeV, respectively. All variables are reconstructed with the SUB method and only the leading Higgs bosons and leading vectorlike leptons are shown.}
\end{figure}

The distributions of the leading Higgs invariant mass, the dimuon stransverse mass, and the transverse momenta of the leading Higgs bosons and leading vectorlike leptons in the SUB Higgs reconstruction method are presented in Fig. \ref{fig:dist} for illustration. We can see that the Higgs boson of the signal with relatively heavy $E^\pm$ can be effectively reconstructed by the SUB method. The invariant masses of fake Higgs jets in the background processes are typically below the true Higgs boson mass. 
The $m_{T2} (\mu_1, \mu_2)$ variable is always larger than $m_H$. In background processes, the hardest constituents are given by the top quark mass, so the distribution of $m_{T2} (\mu_1, \mu_2)$ is cut off at $m_H+m_t$. While in signal processes, the upper bounds on $m_{T2} (\mu_1, \mu_2)$ are given by the masses of the vectorlike leptons, which can be much higher than the top quark mass. 
These features make $m_{T2} (\mu_1, \mu_2)$ very efficient in signal and background discrimination.
Moreover, for a pair of relatively heavy vectorlike leptons, the energy scale of the signal process is much higher than those of background processes, leading to a harder spectrum in the distributions of transverse momenta of Higgs boson and vectorlike leptons.

\subsection{Event selection and signal significance}

The pair production of vectorlike leptons in the model is dominantly given by the $s$-channel $Z/\gamma$ exchanges through electroweak interaction.
The cross section is below $\sim \mathcal{O}(1)$ fb for $m_{E} \gtrsim 1$ TeV at the 14 TeV LHC. A high integrated luminosity would be required to probe the vectorlike leptons with mass around $\mathcal{O}(1)$ TeV scale. 

Our event selections proceed as follows. The preselection requires at least four jets and two opposite sign (OS) muons in the final state. Here, the jets are reconstructed by anti-$k_T$ algorithm with radius parameter $R=0.4$, $p_T>20$ GeV and $|\eta|<2.5$. Three signal regions are defined for selecting the OS dimuon, as given in Table \ref{tab:srs}. Each is suitable for some vectorlike lepton masses. The muons should be within the pseudorapidity region $|\eta|<2.5$.

\begin{table}[htbp]\centering
\begin{tabular}{|c|c|c|c|} \hline
Signal region          & SR1 & SR2 & SR3          \\ \hline
Leading muon        & $p_T>80$ GeV & $p_T>150$ GeV  & $p_T>250$ GeV \\ \hline
Subleading muon & $p_T>30$ GeV & $p_T>80$ GeV & $p_T>150$ GeV \\ \hline
\end{tabular}
\caption{\label{tab:srs}Three signal regions for selecting the OS dimuon.}
\end{table}

Then, we apply a cut on the dimuon stransverse mass, $m_{T2} (\mu_1, \mu_2)>300$ GeV. Furthermore, we require that there are exactly two reconstructed Higgs (either NOR Higgs or SUB Higgs), both of which satisfy $90~\text{GeV}\leq M_H \leq 130~\text{GeV}$. The two Higgs bosons should contain at least three $b$-tagged subjet in total.\footnote{There is no event with both reconstructed NOR Higgs pair and SUB Higgs pair.} In addition, two signal regions are defined for Higgs $p_T$ in Table III, and they are denoted by SR$ij$, where $i=1-3$ stands for signal regions for selecting the OS dimuon, and $j=1,2$ stands for signal regions for selecting the Higgs transverse momentum.
\begin{table}[htbp]\centering
\begin{tabular}{|c|c|c|} \hline
Signal region         & SR$i$1 & SR$i$2          \\ \hline
Leading Higgs boson     &  $p_T>200$ GeV  & $p_T>350$ GeV  \\ \hline
Subleading Higgs boson&  $p_T>150$ GeV & $p_T>250$ GeV  \\ \hline
\end{tabular}
\caption{\label{tab:srs1}Two signal regions for selecting the transverse momenta of Higgs bosons.}
\end{table}

Finally, the SR$ij$ are further divided according to the reconstructed Higgs type and vectorlike lepton mass, as given in Table \ref{tab:srs2}. This gives a total of 24 signal regions in our analysis, i.e., SR$ijk$, $i=1-3$, $j=1,2$, $k=1-4$.  

\begin{table}[htbp]\centering
\begin{tabular}{|c|c|c|c|c|} \hline
Signal region    & SR$ij$1 & SR$ij$2 & SR$ij$3 & SR$ij$4          \\ \hline
Higgs type        &  SUB Higgs  & SUB Higgs  & SUB Higgs & NOR Higgs \\ \hline
Vectorlike lepton & $m_E>350$ GeV & $m_E>450$ GeV & $m_E>650$ GeV & $m_E<300$ GeV \\ \hline
\end{tabular}
\caption{\label{tab:srs2}Four signal regions based on the reconstructed Higgs type and vectorlike lepton mass. The cut is applied on the heavier one of the two reconstructed vectorlike leptons. The index $i$ runs over 1, 2, 3, corresponding to Table \ref{tab:srs}, and the index $j$ runs over 1, 2, corresponding to Table \ref{tab:srs1}. }
\end{table}

Once we obtain the numbers of signal ($s$) and background ($b$) events in each signal region, the signal significance of that signal region can be calculated by \cite{Cowan:2010js}
\begin{equation}
\mathcal{S}=\sqrt{2((s+b)\ln(1+\frac{s}{b})-s)}~.
\end{equation}
For the signal process with given $m_E$, the signal region that provides the highest signal significance is chosen. In Table \ref{tab:bench}, we show the cut flow in the chosen signal regions for three benchmark points. The most sensitive signal regions for $m_{E}=$500, 750, and 1000 GeV are SR212, SR322, and SR323, respectively. 
\begin{table}[htbp]\centering
\begin{tabular}{|c|c|c|c|c|c|c|} \hline
$m_{E}/\text{GeV}$ & 500 & BKG(SR212) & 750 & BKG(SR322) & 1000 & BKG(SR323) \\ \hline
$N_\text{jet}\geq 4~\&~N_{\mu}\geq 2$ & 0.98   & 419  & 0.15   & 61      & 0.32     & 61 \\ \hline
$m_{T2}\geq 300$ GeV                         & 0.44   & 15    & 0.10   & 5.3     & 0.024   & 5.3  \\ \hline
$N_H=2, H_{b\text{-tag}}$                     & 0.073 & 0.16 & 0.020 & 0.040 & 0.0060 & 0.040 \\ \hline
$ M_{E}$ cut                                          & 0.073 & 0.13 & 0.020 & 0.033 & 0.0059 & 0.019 \\ \hline
\end{tabular}
\caption{\label{tab:bench}The cut flow of our analysis for signals (with three representative vectorlike lepton masses 500, 750, and 1000 GeV) and background. The numbers correspond to the production cross sections (in fb) after cuts at the 14 TeV LHC. We have assumed BR$(E \to \mu H) =100\%$.}
\end{table}

\begin{figure}[htbp]
\centering
\includegraphics[width=0.5\textwidth]{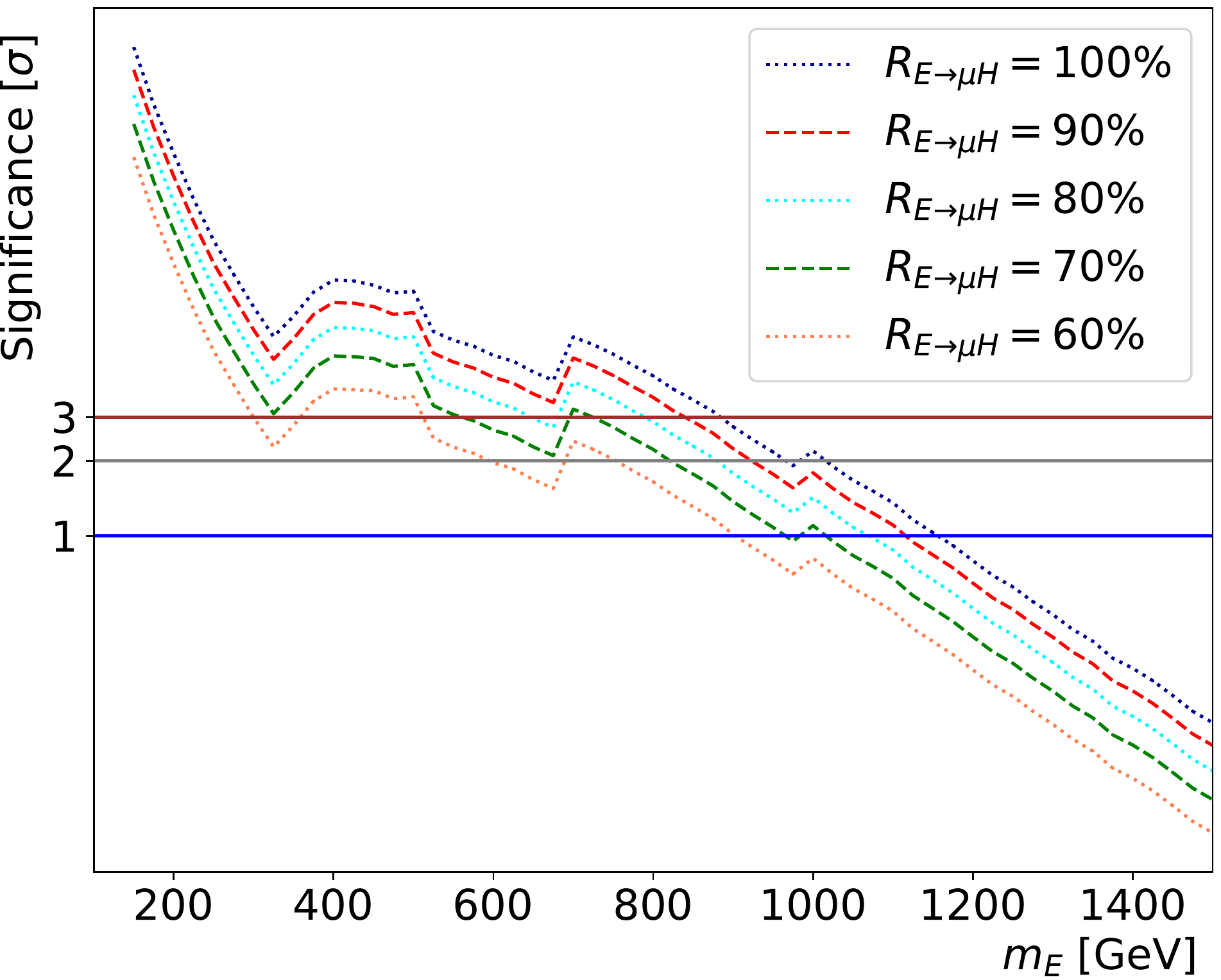}
\caption{\label{fig:sgn} The signal significance of our proposed search for vectorlike leptons with varying mass within [150, 1500] GeV and branching ratio within [60, 100]\%. The analysis is intended for the 14 TeV LHC with an integrated luminosity of 3000 fb$^{-1}$.}
\end{figure}

In Fig. \ref{fig:sgn}, we present the highest signal significance among signal regions with varying vectorlike lepton mass $m_E \in [150,~1500]$ GeV and different branching ratios BR$(E \to \mu H) \in [60,~100]\%$.  It can be seen that the vectorlike lepton mass below $\sim[800,~1000]$ GeV can be probed at 2-$\sigma$ level at the 14 TeV LHC with an integrated luminosity of 3000 fb$^{-1}$.

\section{Summary}\label{sec:sum}

The observed rare $B$-meson decay anomalies might indicate the existence of a $Z'$ boson that has flavor-changing couplings to quarks and nonuniversal couplings to leptons. We considered a U(1)$_X$ extension of the SM gauge group such that the desired types of couplings can be naturally generated by introducing extra vectorlike fermions. 
Taking into account the constraints from $B_s-\bar B_s$ mixing, imposing perturbativity requirements on the NP gauge and Yukawa couplings, the observed $B$-meson anomalies require the masses of new particles in the model to be bounded from above: $m_{Z'} \lesssim 1.8$ TeV, $m_{N,E} \lesssim 9.6$ TeV, and $m_{U,D}\lesssim 77$ TeV.  

The search for $Z'$ in the dimuon final state at the LHC by CMS covered only part of the parameter space. Nonetheless, the couplings of $Z'$ to muon/muon neutrino, the second and third generation quarks are further constrained by the exclusion limits on $Z'$. In terms of branching ratio, for $m_{Z'}<1.5$ TeV, the $Z'$ boson decays into a muon pair or a muon neutrino pair at least 90\% of the time. By extrapolating current exclusion limits to the 14 TeV LHC with an integrated luminosity of 3000 fb$^{-1}$, we find it impractical for future direct search to thoroughly probe a $Z'$ boson with more than 200 GeV, despite the improved sensitivity by an order of magnitude.

On the other hand, the search for the vectorlike lepton is complementary to that for the $Z'$ boson, because its production at the LHC is almost entirely controlled by the SM gauge couplings. In the parameter space where $m_E>m_{Z'}$, the vectorlike lepton decays dominantly into a muon and a $Z'$ boson which subsequently decays into two muons with a certain branching ratio. This gives as much as six muons in the final state. The 6-muon signature is essentially background free, so that a number of events of $\mathcal{O}(10)$ would allow high confidence level signal/exclusion. Our study showed that the 6-muon signature can probe vectorlike lepton mass up to 1400 GeV at the future LHC with an integrated luminosity of 3000 fb$^{-1}$. 
If $Z'$ is heavier than the vectorlike lepton, the $E\to \mu H$ channel will become competitive with or even dominant over the $E \to\mu Z'^{*}(\to \mu \mu)$ channel, especially when the mixing between the SM Higgs and the new scalar field $\phi$ is sizeable. 
We performed a detailed search for the signature of dimuon plus two boosted Higgs bosons from vectorlike lepton pair production. The future LHC is sensitive to the vectorlike lepton with mass below $\sim [800,~1000]$ GeV for BR$(E\to \mu H) \in [60,~100]\%$.
\section*{Acknowledgement}
We thank the Korea Institute for Advanced Study and the Institute of Theoretical Physics, Chinese Academy of Sciences for providing computing resources (KIAS Center for Advanced Computation Linux Cluster System and HPC Cluster of ITP-CAS) for this work.
This research was supported in part by Grants No. 11475238, No. 11647601, and No. 11875062 supported by the National Natural Science Foundation of China, the Key Research Program of Frontier Science, CAS (T. L.), the Fundamental Research Funds for the Central Universities, and the National Research Foundation of Korea (NRF) Grant No. NRF-2015R1A2A1A05001869 (J. L.).

\bibliographystyle{jhep}
\bibliography{ref}

\end{document}